\documentclass[12pt,a4paper]{article}
\usepackage{graphicx}

\begin{document}
\begin{center}
{\Large\bf Generalized Statistics and the Formation of}
\vspace{0.5cm}
 {\Large\bf a Quark-Gluon Plasma}

\vspace{1cm}

A.M. Teweldeberhan$^1$, H.G. Miller$^1$, and R. Tegen$^{1,\:2}$
\vspace{0.5cm}

{\itshape
$^1$Department of Physics, University of Pretoria, Pretoria 0002, South Africa
\vspace{0.5cm}

$^2$Department of Physics, University of the Witwatersrand,  P.O. WITS 2050, 
Johannesburg, South Africa}
\vspace{0.5cm}

September 12, 2002

\vspace{1cm}

{\Large\bf Abstract}
\end{center}
\vspace{0.5cm}

The aim of this paper is to investigate the effect of a non-extensive form of 
statistical mechanics proposed by Tsallis on the formation of a quark-gluon 
plasma (QGP). We suggest to account for the effects of the dominant part of the 
{\itshape long-range} interactions among the constituents in the QGP by a 
change in the statistics of the system in this phase, and we study the 
relevance of this statistics for the phase transition. The results show that 
small deviations ($\approx 10\%$) from Boltzmann-Gibbs statistics in the QGP 
produce a noticeable change in the phase diagram, which can, in principle, be 
tested experimentally.
\newpage

\begin{center}
{\Large\bf 1. Introduction}
\end{center}
\vspace{0.5cm}

In the past few years the non-extensive form of statistical mechanics proposed 
by Tsallis [1] has found applications in astrophysical self-gravitating systems 
[2], solar neutrinos [3], high energy nuclear collisions [4], cosmic microwave 
back ground radiation [5], and high temperature superconductivity [6, 7]. In 
these cases a small deviation of q ($\approx 10\%$)  from Boltzmann-Gibbs (BG) 
statistics reduces the discrepancies between experimental data and theoretical 
models. 

Substantial theoretical research has been carried out to study the phase 
transition between hadronic matter and the QGP [8-12]. When calculating the QGP 
signatures in relativistic nuclear collisions, the distribution functions of 
quarks and gluons are traditionally described by BG statistics. In this paper 
we investigate the effect of the non-extensive form of statistical mechanics on 
the formation of a QGP. The crucial difference between the hadronic and QGP 
phase is the relative importance of {\itshape short-range} and {\itshape 
long-range} interactions among the constituents on either side of the 
anticipated phase transition. The hadronic phase is characterized by a dominant 
{\itshape short-range} interaction among hadrons (which lends itself to BG 
statistics) while the QGP phase has a greatly reduced {\itshape short-range} 
interaction (due to ``asymptotic freedom'') and consequently a dominant 
{\itshape long-range} interaction. We suggest in this communication to account 
for the effects of the dominant part of this {\itshape long-range} interaction 
by a change in statistics for the constituents in the QGP phase. In addition we 
deem the non-dominant part of this {\itshape long-range} interaction negligible 
for the purpose of the phase diagram which we study here in detail. This latter 
view is supported by the empirical insensitivity of the phase diagram to 
{\itshape details} of the interaction among constituents on either side of the 
phase transition. Therefore, we use the generalized statistics of Tsallis to 
describe the QGP phase while maintaining the usual BG statistics in the hadron 
phase (as we shall see this may also be regarded as choosing Tsallis statistics 
in the hadron phase with the Tsallis parameter q=1).

Since hadron-hadron interactions are of {\itshape short-range}, BG statistics 
is successful in describing particle production ratios seen in relativistic 
heavy ion collisions below the phase transition [13-17]. Our motivation for the 
use of generalized statistics in the QGP phase lies in the necessity to include 
the {\itshape long-range} interactions on the QGP side. It has been 
demonstrated [18, 19] that the non-extensive statistics can be considered as 
the natural generalization of the extensive BG statistics in the presence of 
long-range interactions, long-range microscopic memory, or fractal space-time 
constraints. It was suggested in [4] that the extreme conditions of high 
density and temperature in ultrarelativistic heavy ion collisions can lead to 
memory effects and long-range colour interactions. Anticipating the formation 
of a QGP in heavy ion collisions, we use the generalized statistics to 
incorporate the effect of a {\itshape long-range} colour magnetic force in this 
phase. 

The generalized entropy proposed by Tsallis [1] takes the form:
\begin{equation}
S_q=\kappa \frac{(1-\sum_{i=1}^w p_i^q)}{q-1}\hspace{1cm} (q\in\Re),
\end{equation}
where $\kappa$ is a positive constant (from now on set equal to 1), w is the 
total number of microstates in the system, $p_i$ are the associated 
probabilities with $\sum_{i=1}^w {p_i}=1$, and the Tsallis parameter (q) is a 
real number. It is straightforward to verify that the usual BG logarithmic 
entropy, $S=-\sum_{i=1}^w p_i\ln p_i$, is recovered in the limit 
$q\rightarrow1$. Only in this limit is the ensuing statistical mechanics 
extensive [1, 19, 20]. For general values of q, the measure $S_q$ is 
non-extensive. That is, the entropy of a composite system $A\oplus B$ 
consisting of two subsystems A and B, which are statistically independent in 
the sense that $p_{i,j}^{(A\oplus B)}=p_i^{(A)}p_j^{(B)}$, is not equal to the 
sum of the individual entropies associated with each subsystem. Instead, the 
entropy of the composite system is given by Tsallis' q-additive relation [1],
\begin{equation}
S_q(A\oplus B)=S_q(A)+S_q(B)+(1-q)S_q(A)S_q(B) \end{equation}
The quantity $|1-q|$ can be regarded as a measure of the degree of 
non-extensivity exhibited by $S_q$.

The standard quantum mechanical distributions can be obtained from a maximum 
entropy principle based on the entropic measure [21, 22],
\begin{equation}
S=-\sum_i[\bar{n}_i \ln \bar{n}_i\mp (1\pm \bar{n}_i) \ln(1\pm \bar{n}_i)],
\end{equation}
where the upper and lower signs correspond to bosons and fermions, 
respectively, and $\bar{n}_i$ denotes the number of particles in the $i^{th}$ 
energy level with energy $\epsilon_i$. The extremization of the above measure 
under the constraints imposed by the total number of particles,
\begin{equation}
\sum_i \bar{n}_i=N, \end{equation}
and the total energy of the system,
\begin{equation}
\sum_i \bar{n}_i\epsilon_i=E, \end{equation}
leads to the standard quantum distributions,
\begin{equation}
\bar{n}_i=\frac{1}{\exp\beta (\varepsilon_i-\mu)\mp1}, \end{equation}
where $\beta=\frac {1} {T}$ and the upper and lower signs correspond to the 
Bose-Einstein and Fermi-Dirac distributions, respectively.

To deal with non-extensive scenarios (characterized by $q\neq 1$), the extended 
measure of entropy for fermions proposed in [6, 23] is:
\begin{equation}
S_q^{(F)} [{\bar{n}_i}]=\sum_i \{(\frac {\bar{n}_i-\bar{n}_i^q} 
{q-1})+[\frac{(1-\bar{n}_i)-(1-\bar{n}_i)^q)} {q-1}]\},
\end{equation}
which for $q\rightarrow 1$ reduces to the entropic functional (3) (with lower 
signs).

The constraints
\begin{equation}
\sum_i \bar{n}_i^q=N \end{equation}
and
\begin{equation}
\sum_i \bar {n}_i^q \epsilon_i=E \end{equation}
lead to
\begin{equation}
\bar{n}_i=\frac {1} {[1+(q-1)\beta (\epsilon_i-\mu)]^{\frac {1} {q-1}}+1} 
\end{equation}
In the limit $q\rightarrow 1$ one recovers the usual Fermi-Dirac distribution 
(6) (with lower sign).

Similarly,
\begin{equation}
\bar{n}_i=\frac {1} {[1+(q-1)\beta (\epsilon_i-\mu)]^{\frac {1} {q-1}}-1}  
\end{equation}
for bosons. We now turn to the description of the system in the QGP phase.
\vspace{1cm}

\begin{center}
{\Large\bf 2. The QGP phase}
\end{center}
\vspace{0.5cm}

In this approach, the quarks and gluons are treated as forming an ideal gas, 
apart from the non-perturbative corrections to the pressure and energy density 
resulting from the bag model [8]. We initially (a) describe the system by BG 
statistics in order to define our notation and general numerical procedure. In 
(b) we lay out the differences due to the non-extensive statistics. We 
emphasize that only (b) can incorporate the anticipated {\itshape long-range} 
forces in the QGP phase.
\newpage

{\large (a) Boltzmann-Gibbs (BG) Statistics}
\vspace{0.5cm}

According to the BG statistics the energy density, pressure and baryon number 
density for a QGP consisting of massless u and d quarks and anti-quarks, each 
with degeneracy factor $d_Q=12$, and gluons, with degeneracy $d_G=16$, at a 
temperature T and baryon chemical potential $\mu$ are given by \begin{equation}
u_{QGP}=\frac {d_Q} {2\pi^2} \{\int_0^\infty dk \: k^3 \: 
[\bar{n}_Q+\bar{n}_{\bar{Q}}]+\frac {d_G} {2\pi^2} \int_0^\infty dk \: k^3 \: 
\bar{n}_G +B
\end {equation}  \[
p_{QGP}=\frac {d_Q T} {2\pi^2} \int_0^\infty dk \: k^2 \: \{\ln [1+\exp\frac 
{1}{T}(\mu_Q-k)]+\ln [1+\exp\frac {-1}{T}(\mu_Q+k)\}
\]
\begin{equation}
-\frac {d_Q T} {2\pi^2} \int_0^\infty dk \: k^2 \: \ln [1-\exp(\frac {-k} 
{T})]-B
\end{equation}
and
\begin{equation}
n_{QGP}=\frac {d_Q} {6\pi^2} \{\int_0^\infty dk \: k^2 \: [\bar 
{n}_Q-\bar{n}_{\bar {Q}}]
\end{equation}
where
\begin{equation}
\bar{n}_{Q(\bar{Q})}=\frac{1} {\exp \frac{1} {T} (k\mp\mu_Q)+1},
\end{equation}
\begin{equation}
\bar{n}_G=\frac{1} {\exp (\frac{k} {T})-1}
\end{equation}
and $\mu_Q=\frac {\mu} {3}$. 

Integration by parts of equation (13) yields
\begin{equation}
p_{QGP}=\frac {1} {3} (u_{QGP}-4B)
\end{equation}

Integration of equations (12)-(14) yields
\begin{equation}
u_{QGP}=\frac {\pi^2} {30} (d_G+\frac {7}{4} d_Q) T^4+\frac {d_Q \mu^2 T^2} 
{36}+\frac {d_Q \mu^4} {648\pi^2}+B \end{equation}
\begin{equation}
p_{QGP}=\frac {\pi^2} {90} (d_G+\frac {7}{4} d_Q) T^4+\frac {d_Q \mu^2 T^2} 
{108}+\frac {d_Q \mu^4} {1944\pi^2}-B \end{equation}
and
\begin{equation}
n_{QGP}=d_Q[\frac {\mu T^2} {54}+\frac {\mu^3} {486 \pi^2}] \end{equation}
where B is the bag constant which is taken here as (210 MeV)$^4$ [9] with an 
uncertainty of $\approx 15\%$.
\vspace{1cm}

{\large (b) Generalized Statistics}
\vspace{0.5cm}

If we use the generalized statistics to describe the entropic measure of the 
whole system, the distribution function can not , in general, be reduced to a 
finite, closed, analytical expression [23-28]. For this reason we use 
generalized statistics to describe the entropies of the individual particles, 
rather than of the system as a whole\footnote{For a more detailed account of 
this important point see ref. [23].}.

The distribution function is then given by \begin{equation}
\bar{n}_i=\frac {1} {[1+(q-1)\beta (\epsilon_i-\mu)]^{\frac {1} {q-1}}\mp1},  
\end{equation}
where the upper and lower signs correspond to bosons and fermions, respectively 
[23]. In this case
\begin{equation}
\bar{n}_{Q(\bar{Q})}=\frac {1} {[1+\frac{1} {T}(q-1)(k\mp\mu_Q)]^{\frac {1} 
{q-1}}+1}
\end{equation}
and
\begin{equation}
\bar{n}_G=\frac {1} {[1+\frac{1} {T}(q-1)k]^{\frac {1} {q-1}}-1}
\end{equation}

In the limit $q\rightarrow 1$ one recovers equations (15) and (16).

The expression for the pressure is given by
\[
p_{QGP}=\frac {d_Q T} {2\pi^2} \int_0^\infty dk \: k^2 \: (\frac{f_Q^{q-1}-1} 
{q-1}+\frac{f_{\bar{Q}}^{q-1}-1} {q-1})-
\]
\begin{equation}
\frac {d_G T} {2\pi^2} \int_0^\infty dk \: k^2 \: (\frac{f_G^{q-1}-1} {q-1})-B
\end{equation}
where
\begin{equation}
f_Q=1+[1+\frac {1}{T}(q-1)(k-\mu_Q)]^\frac{1}{1-q}
\end{equation} \begin{equation}
f_{\bar{Q}}=1+[1+\frac {1}{T}(q-1)(k+\mu_Q)]^\frac{1}{1-q}
\end{equation}
and
\begin{equation}
f_G=1-[1+\frac{1}{T}(q-1)k]^\frac{1}{1-q}
\end{equation}
which in the limit $q\rightarrow 1$ reduces to (13).

Since the integrals in (12)-(14) are not integrable analytically one has to 
calculate these integrals numerically. For $q>1$, the quantity 
$[1+\frac{1}{T}(q-1)(k-\mu_Q)]$ becomes negative if $\mu_Q>k$. To avoid this 
problem we use [29],
\begin{equation}
f_Q=1+[1+\frac {1}{T}(q-1)(k-\mu_Q)]^\frac{1}{1-q},\hspace{1cm} k\geq\mu_Q
\end{equation}
and
\begin{equation}
f_Q=1+[1+\frac {1}{T}(1-q)(k-\mu_Q)]^\frac{1}{q-1},\hspace{1cm} k<\mu_Q
\end{equation}
In the limit $q\rightarrow 1$ one recovers, of course, the appropriate 
Fermi-Dirac distribution in both cases. We now turn to the hadronic phase of 
the system which is more readily accessible to experiment.
\vspace{1cm}

\begin{center}
{\Large\bf 3. The hadron phase}
\end{center}
\vspace{0.5cm}

The hadron phase is taken to contain only interacting nucleons and antinucleons 
and an ideal gas of massless pions motivated by the findings in [30]. The 
interactions between nucleons can be treated either by means of an excluded 
volume approximation or by a mean field approximation. The excluded volume 
approximation to hadron interactions is either thermodynamically inconsistent, 
or leads to computational difficulties in realistic calculations if the 
consistency problem is addressed. To treat the interactions we use the 
relativistic mean field or Hartree approximation proposed by Walecka [31] which 
is thermodynamically self-consistent. In this model the interaction between 
nucleons is described by the scalar-isoscalar $\sigma$ and vector-isoscalar 
$\omega^\mu$ mesonic fields with baryon-meson interaction terms in the 
Lagrangian: $g_\sigma\bar{\psi}\psi\sigma$ and 
$g_\omega\bar{\psi}\gamma^\mu\psi \omega_\mu$. For nuclear matter in 
thermodynamical equilibrium these mesonic fields ($\sigma$ and $\omega^\mu$) 
are considered to be constant classical quantities. The scalar field $\sigma$ 
describes the attraction between nucleons and lowers the nucleon (antinucleon) 
mass M to $M^*=M-g_\sigma\langle\sigma\rangle$. The nucleon-nucleon repulsion 
is described by the vector field $\omega^\mu$ which changes the nucleon 
(antinucleon) energy by $(\pm U(n))$ \footnote{The odd G-parity of the 
$\omega$-meson is responsible for the attractive $\omega$-exchange in $N\bar 
{N}$ scattering as compared to the repulsive $\omega$-exchange in 
$NN(\bar{N}\bar{N})$ scattering.}.

The thermodynamically self-consistent equations of state (EOS) for nuclear 
matter are [32, 33]:
\[
p(T, \mu)=\frac{\gamma_N} {3}\int \frac {d^3k} {(2\pi)^3} \frac {k^2} 
{\sqrt{k^2+{M^*}^2}} (\bar{n}_N+\bar{n}_{\bar{N}})+ n U(n)-
\]
\begin{equation}
\int_0^n dn' U(n')+P(M^*) \end{equation}
\begin{equation}
\bar{n}_{N(\bar{N})}=[\exp (\frac {\sqrt{k^2+{M^*}^2}\mp\mu\pm U(n)} 
{T})+1]^{-1} \end{equation}
\begin{equation}
(\frac {\delta P} {\delta M^*})_{T, \mu}\equiv \frac {dP(M^*)} {dM^*}-\gamma_N 
\int \frac {d^3k} {(2\pi)^3} \frac {M^*} {\sqrt{k^2+{M^*}^2}} 
(\bar{n}_N+\bar{n}_{\bar{N}})=0 \end{equation}
\begin{equation}
n(T, \mu)=\gamma_N \int \frac {d^3k} {(2\pi)^3} (\bar{n}_N-\bar{n}_{\bar{N}})
\end{equation}
and
\begin{equation}
u(T, \mu)=\gamma_N \int \frac {d^3k} {(2\pi)^3}\sqrt{k^2+{M^*}^2} 
(\bar{n}_N+\bar{n}_{\bar{N}})+\int_0^n dn' U(n')-P(M^*),  \end{equation}
where p, n, and  u are the pressure, baryon number density, and energy density 
respectively, $\bar{n}_{N(\bar{N})}$ is the distribution function of nucleons 
(antinucleons), $\mu$ is the baryon chemical potential, and $\gamma_N$ is the 
spin-isospin degeneracy of the nucleon which is 4 for symmetric nuclear matter. 
Equation (32) describes the dependence of the effective nuclear mass $M^*$ on T 
and $\mu$ which is defined by extremizing the thermodynamical potential 
(maximizing the pressure).

If we choose [31], \begin{equation}
P(M^*)=-\frac {1} {2C_\sigma^2} (M-M^*)^2,\hspace{1cm} U(n)=C_\omega^2n
\end{equation}
where $C_\sigma\equiv g_\sigma (\frac {M} {m_\sigma})$, and $C_\omega\equiv 
g_\omega (\frac {M} {m_\omega})$. The parameter set (M, $g_\sigma$, $g_\omega$, 
$m_\sigma$, $m_\omega$) consists of phenomenological constants determined by 
experiment. We take here M= 0.940 GeV, $M^*$= 0.543M, $g_\sigma\approx 11$ 
GeV$^{-1}$,  $g_\omega\approx 14$ GeV$^{-1}$, $m_\sigma$= 0.520 GeV, 
$m_\omega$= 0.783 GeV, which reproduces data (see ref. [33, 34] for details) in 
the hadronic phase.

The energy density, pressure, and baryon number density for the hadron gas 
taken to contain nucleons, antinucleons and an ideal gas of massless pions are:
\[
u_H(T, \mu)=\frac {\gamma_N} {2\pi^2} \int_0^\infty dk \; k^2 
\sqrt{k^2+{M^*}^2} (\bar{n}_N+\bar{n}_{\bar{N}})+\frac {1} {2} C_\omega^2n^2+
\]
\begin{equation}
\frac{1} {2C_\sigma^2} (M-M^*)^2 +\frac {1} {10} \pi^2 T^4 \end{equation}
\[
p_H(T, \mu)=\frac {\gamma_N} {6\pi^2} \int_0^\infty \frac {dk \: k^4} 
{\sqrt{k^2+{M^*}^2}} (\bar{n}_N+\bar{n}_{\bar{N}})+\frac {1} {2} C_\omega^2n^2-
\]
\begin{equation}
\frac{1} {2C_\sigma^2} (M-M^*)^2+\frac {1} {30} \pi^2 T^4   \end{equation}
and
\begin{equation}
n_H(T, \mu)=\frac {\gamma_N} {2\pi^2} \int_0^\infty dk \: k^2  
(\bar{n}_N-\bar{n}_{\bar{N}}),
\end{equation}
where $\bar{n}_{N(\bar{N})}$ as in (31) with U(n) as in (35).

Finally, we address the phase transition from the hadronic to the QGP phase, 
within our model.

\newpage

\begin{center}
{\Large\bf 4. Phase transition}
\end{center}
\vspace{0.5cm}

Assuming a first order phase transition between hadronic matter and QGP one 
matches an EOS for the hadronic system and the QGP via Gibbs conditions for 
phase equilibrium:
\begin{equation}
p_H=p_{QGP},\hspace{1cm} T_H=T_{QGP}, \hspace{1cm} \mu_H=\mu_{QGP}
\end{equation}
With these conditions the pertinent regions of temperature T and baryon 
chemical potential $\mu$  are shown in fig. 1 for q= 1, 1.1 (0.9) and 1.25 
(0.75). The critical temperature at $\mu=0$ for q= 1, 1.1 (0.9) and 1.25 (0.75) 
are found to be 148 MeV, 122 MeV and 79 MeV, respectively. As the non-extensive 
parameter deviates from q= 1 to 1.25 (0.75), the critical temperature becomes 
almost independent of the baryon chemical potential which is associated with 
the number of particles. The variation of the bag constant B between (180 
MeV)$^4$ and (250 MeV)$^4$ does not alter our findings significantly. One still 
observes a flattening of the T($\mu$) curves in fig. 1 as $|1-q|$ increases. 
The only effect is a shift of the value of the maximal $\mu$ and T(0) in fig. 
1. There is no significant change of the slope of T($\mu$) over the depicted 
range. The agreement between BG and the generalized statistics as the critical 
temperature approaches zero is evident from equations (13) and (24). Fig. 2 
shows the dependence of the critical temperature on the non-extensive parameter 
at $\mu=0$, 250 MeV and 1000 MeV. The dependence is almost linear for small 
values of q with a slope $|\frac {\Delta T} {\Delta q}|\approx$ 240 MeV. This 
can be interpreted as a new type of universality condition [6] which suggests 
that the formation of a QGP occurs (almost) independent of the total number of 
baryons in heavy ion collisions.
\newpage

\begin{center}
{\Large\bf 5. Conclusion}
\end{center}
\vspace{0.5cm}

We have studied the phase transition from a system in the hadronic phase to the 
QGP phase. On the hadronic side the detailed form of the ({\itshape 
short-range}) interactions among hadrons turns out to be unimportant for the 
phase transition. On the QGP side the {\itshape short-range} interactions have 
died out (due to ``asymptotic freedom'') and have made room for the only 
remaining {\itshape long-range} interactions among the constituents. We take 
account of the dominant part of this interaction by a change in the statistics 
of the system in the QGP phase. We present here testable consequences of using 
the non-extensive form of statistical mechanics proposed by Tsallis in the QGP. 
The resulting insensitivity of the critical temperature to the total number of 
baryons presents a clear experimental signature for the existence of 
non-extensive statistics for the constituents of the QGP.

\vspace{5cm}

\begin{figure}[h] \centering
\begin{center}
\includegraphics [scale=0.5, angle=0]{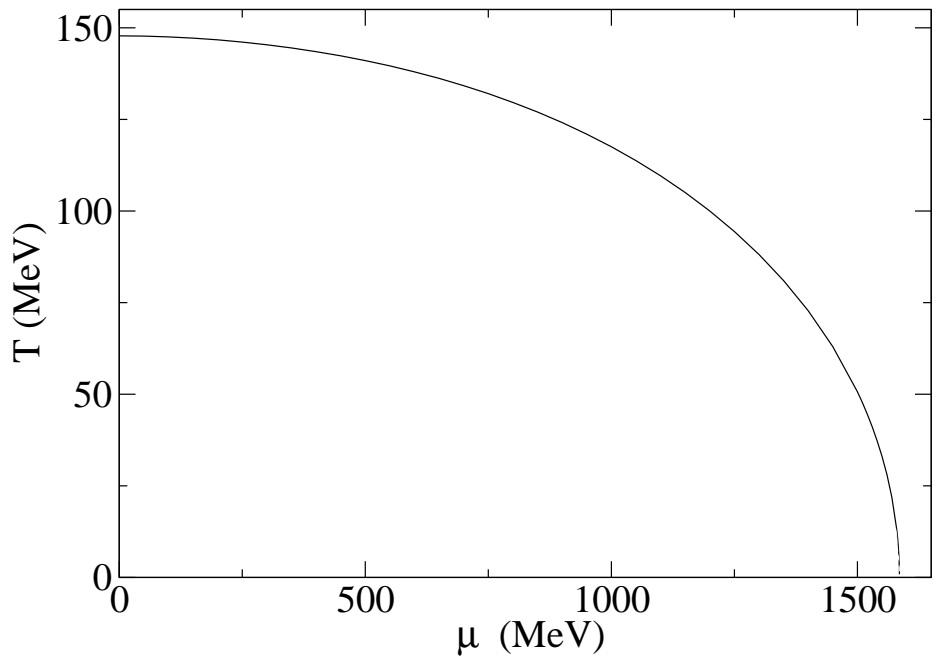}
\end{center}
\caption{Phase transition curves between the hadronic matter and QGP for q=1 
(solid line), q=1.1 (0.9) (dotted line) and q=1.25 (0.75) (dashed line).}
\end{figure}
\vspace{5cm}

\begin {figure}[h] \centering
\begin{center}
\includegraphics [scale=0.5, angle=0]{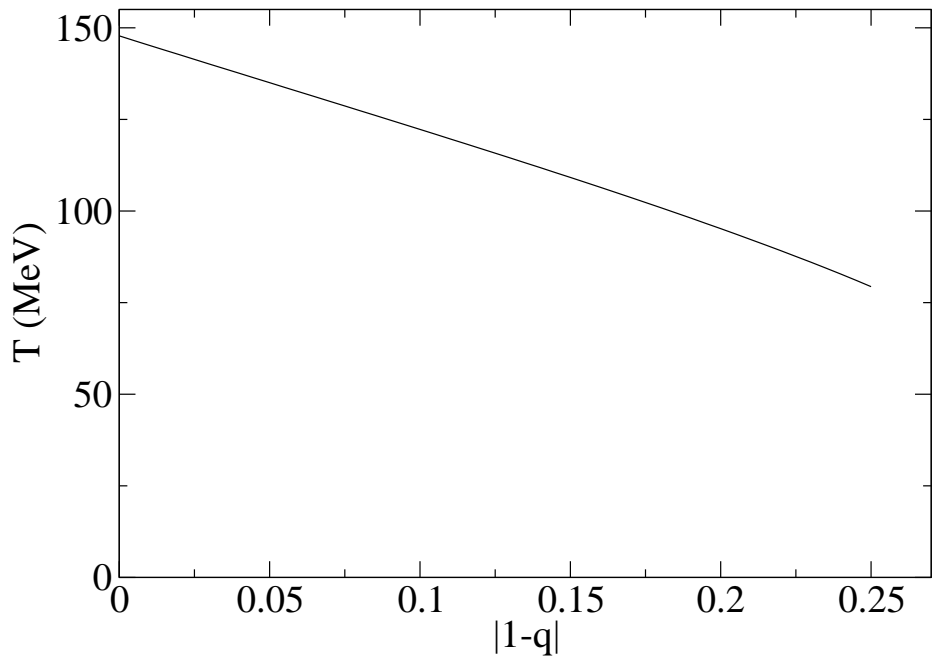}
\end{center}
\caption {The dependence of T on $|1-q|$ at $\mu$= 0 (solid line), $\mu$= 250 
MeV (dotted line) and $\mu$= 1000 MeV (dashed line).}
\end{figure}

\end{document}